\documentclass[aps,twocolumn,floatfix,superscriptaddress,longbibliography]{revtex4-1}

\usepackage{graphicx}
\usepackage{indentfirst}
\usepackage{latexsym}
\usepackage{multirow}
\usepackage{tabls}
\usepackage{epsfig}
\usepackage{multirow}
\usepackage{subfigure}
\usepackage{comment}
\usepackage{hyperref}

\usepackage{color}
\usepackage[usenames,dvipsnames]{xcolor}
\usepackage{amssymb}

\usepackage{amsfonts}
\usepackage{amsmath}
\usepackage{bm}
\usepackage{mathrsfs}

\usepackage{algorithm}
\usepackage{algpseudocode} 
\usepackage{setspace} 

\newcommand{\beq}{\begin{equation}}
\newcommand{\eeq}{\end{equation}}
\newcommand{\bea}{\begin{eqnarray}}
\newcommand{\eea}{\end{eqnarray}}
\newcommand{\ba}{\begin{array}}
\newcommand{\ea}{\end{array}}

\begin{document}

\title{
Accelerated gravitational-wave parameter estimation with reduced order modeling
}

\def\addCambridge{Institute of Astronomy, Madingley Road, Cambridge, CB30HA, United Kingdom}
\def\addDAMTP{Department of Applied Mathematics and Theoretical Physics, Wilberforce Road, Cambridge CB3 0WA, United Kingdom}

\author{Priscilla Canizares}
\affiliation{\addCambridge}\affiliation{\addDAMTP}

\def\addUMDb{Department of Physics, Joint Space Sciences Institute, Maryland Center for Fundamental Physics. University of Maryland, College Park, MD 20742, USA}
\def\addCornell{Center for Radiophysics and Space Research, Cornell University, Ithaca, New York 14853, USA}
\author{Scott E. Field}
\affiliation{\addUMDb}\affiliation{\addCornell}

\author{Jonathan Gair}
\affiliation{\addCambridge} 

\def\addLIGO{LIGO, California Institute of Technology, Pasadena, CA 91125, USA}
\author{Vivien Raymond}
\affiliation{\addLIGO}

\author{Rory Smith}
\affiliation{\addLIGO}

\def\addUCSD{Center for Astrophysics and Space Sciences,
Center for Computational Mathematics, 
San Diego Supercomputer Center, 
University of California San Diego, 
9500 Gilman Drive, La Jolla, CA 92093-0424}
\author{Manuel Tiglio}
\affiliation{\addUCSD}

\date\today

\begin{abstract}
Inferring the astrophysical parameters of coalescing compact binaries is a key science goal of the upcoming advanced LIGO-Virgo gravitational-wave detector network and, more generally, gravitational-wave astronomy. However, current approaches to parameter estimation for these detectors require computationally expensive algorithms. Therefore there is a pressing need for new, fast and accurate Bayesian inference techniques. In this Letter we demonstrate that a reduced order modeling approach enables rapid parameter estimation to be performed. By implementing a reduced order quadrature scheme within the LIGO Algorithm Library, we show that Bayesian inference on the $9$-dimensional parameter space of non-spinning binary neutron star inspirals can be sped up by a factor of $\sim 30$ for the early advanced detectors' configurations (with sensitivities down to around $40\,$Hz) and $\sim 70$ for sensitivities down to around $20\,$Hz. This speed-up will increase to about $150$ as the detectors improve their low-frequency limit to $10$Hz, reducing to hours analyses which could otherwise take months to complete. Although these results focus on interferometric gravitational wave detectors, the techniques are broadly applicable to any experiment where fast Bayesian analysis is desirable. 
\end{abstract}


\maketitle

\textit{Introduction--}
Advanced LIGO (aLIGO) \cite{Harry2010} and advanced Virgo (AdV) \cite{aVirgo} are expected to yield the first direct detections of gravitational waves (GWs) from astrophysical sources in the next few years. Compact binary coalescences (CBCs) are the most promising GW sources, with expected detection rates between a few and tens per year \cite{Abadie:2010cf}. Effective parameter estimation for CBCs has been demonstrated~\cite{Raymond:2009,Smith:2012du,LVCpe:2013}, but approaches to date carry high computational costs for the cases of interest, even when using efficient algorithms such as Markov chain Monte Carlo (MCMC) or nested sampling~\cite{Skilling:2004}. For the advanced detectors, which will start taking data within a year or two, current approaches will lead to months or years of computational wall (clock) time for the analysis of  each detected signal. Therefore, given the expected detection rates, new techniques which can estimate the astrophysical source parameters in short timescales are highly desirable. Such techniques are also important 
 for large scale mock data challenges.

In parameter estimation studies,  the posterior probability density function (PDF)  of a set of parameters, $\vec{\theta}$, is computed from a GW model, $h(\vec{\theta})$,  assumed to describe the detector's signal $d$. The PDF is related to the likelihood function, $\mathcal{L}(d | \vec{\theta})$, and the prior probability on the model parameters, $\mathcal{P}( \vec{\theta} )$, via Bayes' theorem: $p(\vec{\theta}| d)  \propto \mathcal{P}( \vec{\theta})\ \mathcal{L}(d | \vec{\theta})$. 

Assuming that the detector data $d$ contains the source's signal $h(\vec{\theta}_{\tt true})$ and stationary Gaussian noise $n$, the likelihood function is  given by 
\beq
\log\mathcal{L}(d|\vec{\theta}) = (d|h(\vec{\theta})) - \frac{1}{2} \left[ (h(\vec{\theta})|h(\vec{\theta})) + (d|d) \right] \, ,  \label{eq:likelihood}
\eeq
where $d = h(\vec{\theta}_{\tt true}) +  n$ and $(a | b)$ is a weighted inner product for discretely sampled noisy data 
\beq
(d|h(\vec{\theta})) = 4\Re\ \Delta f \sum_{k=1}^{L} \frac{\tilde d^*(f_k)\tilde h(\vec{\theta};f_k)}{S_{n}(f_k)}\, , \label{e:inner}
\eeq
where $\tilde d(f_k)$ and $\tilde h(\vec{\theta};f_k)$ are the discrete Fourier transforms at frequencies $\{ f_k \}_{k=1}^L$ (with units of Hz), $^{*}$ denotes complex conjugation, and the power spectral density (PSD)  $S_{n}(f_k)$ characterizes the detector's noise. 

For a given observation time $T=1/\Delta f$ and detection frequency window $(f_{\tt high} - f_{\tt low})$ there are 
\beq
L = 
{\tt int}\left( \left[ f_{\tt high} - f_{\tt low} \right] T \right) \label{eq:Ldef}
\eeq
sampling points in the sum (\ref{e:inner}).
When $L$ is large, as in the cases of interest for this Letter, there are two major bottlenecks: $(i)$ evaluation of the model at each $f_k$ and, $(ii)$ assembly of the likelihood (\ref{eq:likelihood}). 

In general, smoothly parameterized models 
are amenable to dimensional reduction which, in turn, provides computationally efficient representations. The specific application of dimensional reduction we consider in this Letter tackles the two aforementioned bottlenecks by permitting the inner product \eqref{e:inner} to be computed with significantly fewer terms. In summary: if a reduced set of $N < L$ basis can be found which accurately spans the model space, it is possible to replace the 
inner product~(\ref{e:inner}) with a reduced order quadrature (ROQ) rule~(\ref{e:roq}) 
containing only $N$ terms, reducing the overall parameter estimation analysis cost by a factor of $L/N$, provided the waveforms can be directly evaluated. For other models, in particular those described by partial or ordinary differential equations, direct evaluation may be accomplished using surrogates~\cite{Field:2013cfa,Purrer:2014fza}.

In this Letter, we demonstrate a ROQ accelerated GW parameter estimation study. While the approach is applicable to any GW model, here we focus on binary neutron star (BNS) inspirals, as these are expected to have the highest detection rates with the lowest uncertainty~\cite{Abadie:2010cf}. We show, both through operation counts and an implementation in the LIGO Algorithm Library (LAL) pipeline \cite{LAL_software}, that ROQs provide a factor of $\sim30$ speedup for the early advanced detectors' configuration \cite{2013arXiv1304.0670L} and $\sim 70$ as the detectors' low-frequency sensitivity reaches $20\,$Hz. This speedup will rise to $\sim150$ as the sensitivity band is lowered to a target of $10$Hz, allowing for significant reduction of the computational cost of Bayesian parameter estimation analyses on BNS sources.  

\textit{Compressed likelihood evaluations--} 
Compared to previous work on which this Letter is based~\cite{antil2012two,Canizares:2013ywa}, parameter estimation of gravitational waves from binary neutron stars carries a number of challenges unique to large datasets which contain long gravitational waveforms with many in-band wave cycles. We briefly summarize  the construction of ROQs while focusing on technical but essential solutions to these challenges. 

Reduced order quadratures can be used for fast parameter estimation whenever the waveform model is amenable to dimensional reduction, through  three steps. The first two are carried out offline, while the final, data-dependent step is performed at the beginning of the parameter estimation analysis. 
(1) Construct a reduced basis, i.e. a set of $N$ elements whose span reproduces the GW model within a specified precision. (2) Construct an empirical interpolant by requiring it to exactly match any template at  
$N$ carefully chosen frequency subsamples $\{F_k\}_{k=1}^N$ \cite{Barrault2004667,Maday_2009,chaturantabut:2737}. (3) The empirical interpolant is used to replace, without loss of accuracy, inner product evaluations~\eqref{e:inner} by ROQ compressed ones~\eqref{e:roq}.\\

\noindent {\it Step 1.} 
The reduced basis set only needs to be built over the space of intrinsic parameters for the waveform family. Furthermore, if the basis is generated using a PSD of unity the representation of the waveform family can be used with any PSD whenever the weights are built as in Eq.~\eqref{eq:wk}.
 
Basis generation in this Letter proceeds in two stages. A greedy algorithm first identifies a preliminary basis suitable for any value of $\Delta f$~\cite{PhysRevD.86.084046}.
Next, this preliminary basis is evaluated at $L$ equally spaced frequency samples appropriate for the detector. The elements of this ``resampled basis" are neither orthogonal nor linearly independent, and so a second similar dimensional reduction is necessary. During these steps appropriately conditioned numerical algorithms ~\cite{Hoffmann_IMGS} are used to avoid poor conditioning, which, for large values of $L$, would otherwise lead to bases with no accuracy whatsoever. \\

\noindent {\it Step 2.}  Given an $N$-size basis it is possible to uniquely and accurately reconstruct any waveform from only $N$ evaluations $\{\tilde{h}(\vec{\theta};F_k)\}_{k=1}^N$. The special frequencies $\{F_k\}_{k=1}^N$, selected from the full set $\{f_i\}_{i=1}^L$, can be found from Algorithm~5 of Ref.~\cite{antil2012two} without modification. This step provides a near-optimal compression strategy in frequency which is complimentary to the parameter one of Step (1). The model's empirical interpolant, valid for all parameters, can be written as (cf. Eq.~(19) of Ref.~\cite{Field:2013cfa})
\beq
\tilde{h}(\vec{\theta}; f_i) \approx \mathrm{e}^{-2 \pi \mathrm{i} t_{c} f_i } \sum_{j=1}^N B_j (f_i) \tilde{h}(\vec{\theta},t_c=0; F_j) \, ,
\label{eq:EIM}
\eeq 
a sum over the basis set $\{ B_j \}_{j=1}^N$ and where, for the sake of the discussion below, we have temporarily isolated the coalescence time $t_{c}$ from the other
parameters. \\

\noindent {\it Step 3.} All extrinsic parameters, except the coalescence time $t_c$, do not affect the frequency evolution of the binary and 
simply scale the inner product~\eqref{e:inner}, thereby sharing the same ROQs. The coalescence time, however, requires special treatment. Substituting Eq.~\eqref{eq:EIM}  into Eq.~(\ref{e:inner}), 
\beq
(d|h(\vec{\theta},t_{c})) 
= \sum_{k=1}^N \omega_k(t_{c}) \tilde{h}(\vec{\theta},t_c=0; F_k) \, ,
\label{e:roq}
\eeq
with the ROQ weights given by 
\beq
\omega_k(t_{c}) = 4\Re\ \Delta f \sum_{i=1}^L \frac{ \tilde{d}^*(f_i) B_k (f_i)}{ S_n(f_i) } \mathrm{e}^{-2 \pi \mathrm{i} t_{c} f_i } \, .
\label{eq:wk}
\eeq

Our approach for the dependence of (\ref{eq:wk}) on $t_c$ is through domain decomposition: an estimate for the time window $W$ centered around the coalescence time $t_{\tt trigger}$ is given by the GW search pipeline. This suggests a prior interval $[ t_{\tt trigger} - W, t_{\tt trigger} + W] $ be used for $t_c$. The prior interval is then split into $n_c$ equal subintervals
of size $\Delta t_c$. The number of subintervals is chosen so that the discretization error is below the measurement uncertainty on the coalescence time.  Finally, on each subinterval a unique set of ROQ weights is constructed. 

Since Step (3) is currently implemented in the LAL pipeline, we summarize it in Algorithm~(\ref{alg:Weights}). The offline steps (1) and (2) are carried out independently. Our approach guarantees, though, that those steps need be to carried out only once for each waveform model. \\

To quickly compute the likelihood we also need an inexpensive rule for $(h(\vec{\theta})|h(\vec{\theta}))$, whose evaluation no longer depends on the data stream or coalescence time. Consequently, such expressions are typically simple. For example, the norm of the restricted TaylorF2 gravitational waveform model considered below is exactly computable \cite{PhysRevD.85.122006}. In the general case, building a basis for $\{ h^2 \}$ and an associated ROQ provides fast norm evaluations. The TaylorF2 gravitational waveform model's norm is computable as a $1$-term ROQ rule, for example.

By design, weight generation is computed in the startup stage for each detection-triggered data set, requiring $N$ full inner product (\ref{e:inner}) evaluations for each $t_c$ interval. This cost is negligible, while each likelihood is subsequently calculated millions of times, leading to significant speedups in parameter estimation studies. The latter scales as the fractional reduction $L/N$  of the number of terms in the quadrature rules (\ref{e:inner}) and (\ref{e:roq}). 

\textit{Parameter estimation acceleration for binary neutron star signals --}  
The majority of a binary neutron star's GW signal will be in the inspiral regime \cite{lrr-2009-2}, which can be described by the closed-form TaylorF2 approximation~\cite{Buonanno:2009zt}. While TaylorF2 does not incorporate spins or the merger-ringdown phases of the binary's evolution, these should not be important for BNS parameter estimation and can therefore be neglected~\cite{Singer:2014qca}. Even for this simple to evaluate waveform family, inference on a single data set requires significant computational wall-time with standard parameter estimation methods~\cite{LVCpe:2013}. We now report on the anticipated speedup $L/N$ achieved by ROQ compressed likelihood evaluations. First, we compute the 
observation time $T$ required to contain a typical BNS signal. Next, we find the number of reduced basis elements $N$ needed to represent this model for any pair of BNS masses. In our studies we fix $f_{\tt high}$ to $1024$Hz while $f_{\tt low}$ varies between $10$Hz and $40$Hz.

\begin{figure}[ht]
\includegraphics[width=0.98\linewidth]{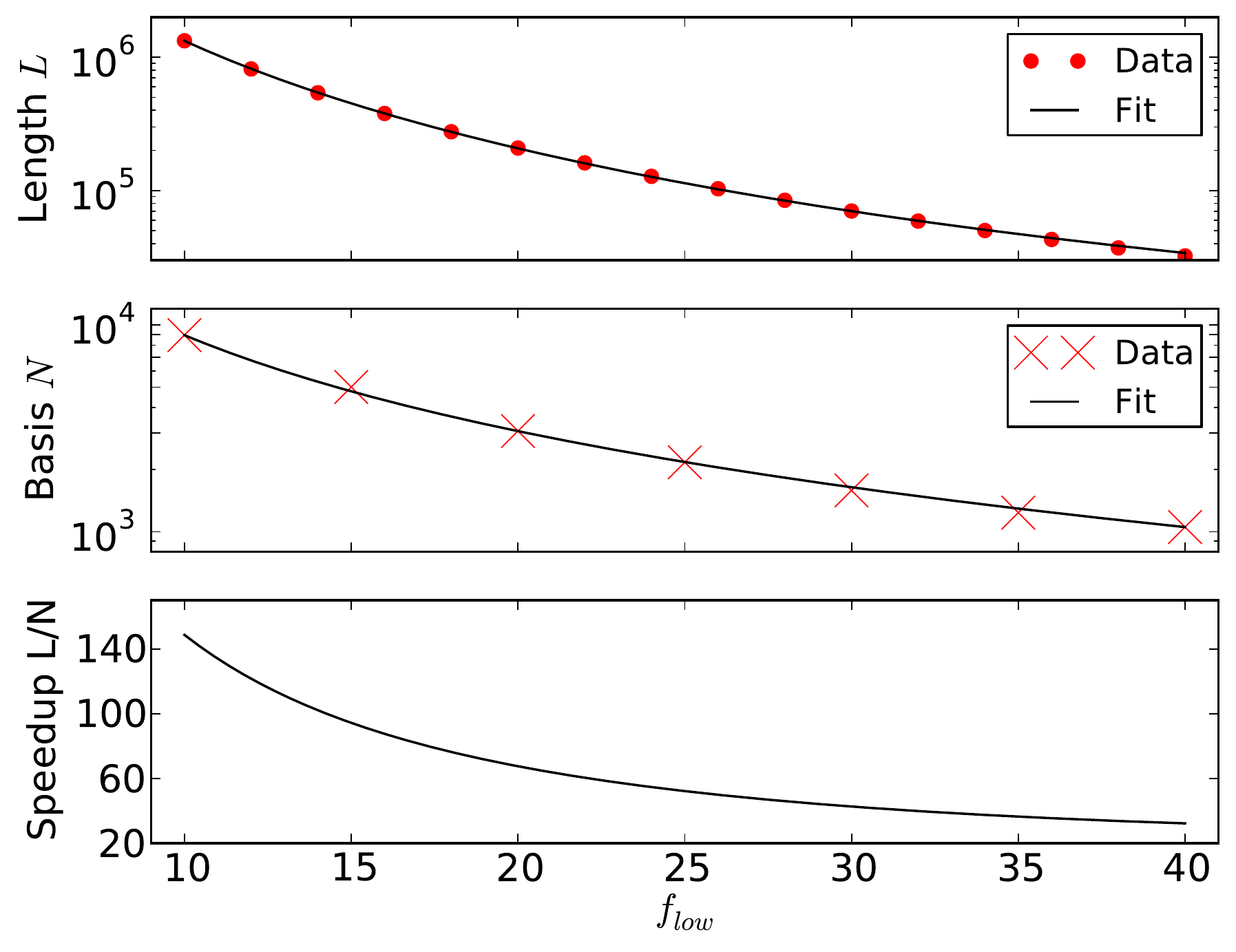}
\caption{{\bf Top}: Length $L$ (red dots) of a typical   
binary neutron star inspiral waveform, with the solid black curve connecting this data implied by the fit~(\ref{eq:fitL}). {\bf Middle}: Number of reduced basis waveforms (red crosses), with the solid black curve given by the fit~(\ref{eq:fitNrb}). {\bf Bottom}: Speedup implied by operation counts, as given by equation~(\ref{eq:Cr_spa}).}
\label{fig:Nbasis_speedup}
\end{figure}

The time taken for a BNS system with an initial GW frequency of $f_{\tt low}$ to inspiral to
$1024$Hz,
\begin{align}
\label{eq:fitL}
T_{\tt BNS} =  \left[ 6.32 + 2.07\times \frac{10^6}{ \left( f_{\tt low}/\text{Hz} \right)^3 + 5.86\left( f_{\tt low}/\text{Hz} \right) ^2  } \right ] {\text s}\, ,
\end{align}
is empirically found by generating a $\left(1+1\right)\,M_{\odot}$ waveform (directly given in the frequency domain) and Fourier transforming to the time domain where we measure the duration up to when the waveform's evolution terminates.
Equation~\eqref{eq:fitL} and subsequent fits were found using a genetic algorithm-based symbolic regression software, {\tt Eureqa}~\cite{eureqa,Schmidt03042009}. The length $L$, as implied by Eq.~(\ref{eq:Ldef}), is plotted in the top panel of Fig.~\ref{fig:Nbasis_speedup}.

As discussed, each basis only needs to be constructed over the space of intrinsic parameters --- in this case the two-dimensional space of 
component masses in the range $\left[1,4\right]M_{\odot}$.  This range is wider than expected for neutron stars, but ensures that the resulting PDFs do not have sharp cut-offs \cite{Mandel:2014tca}. The number of reduced basis required to represent the TaylorF2 model within this range with a representation error around double precision ($\sim 10^{-14}$) can be fit by 
\begin{align}
\label{eq:fitNrb}
N_{\tt BNS} = 3.12\times 10^5 \left( f_{\tt low}/\text{Hz} \right)^{-1.543} \, , 
\end{align}
and is depicted in the middle panel of Fig.~\ref{fig:Nbasis_speedup}.We have found that increasing the high-frequency cutoff to
$4096$ Hz only adds a handful of basis elements, while $L$ changes by a factor of $4$, thus indicating that the speedup for an inspiral-merger-ringdown model might be higher, especially given that not many empirical interpolation nodes are needed for the merger and ringdown regimes \cite{Field:2013cfa}. 

Recalling equation (\ref{eq:Ldef}), the speedup from standard to ROQ-compressed likelihood evaluations is given by 
\begin{align}
\label{eq:Cr_spa} 
\frac{L}{N} \approx \left( 1024 \text{Hz} - f_{\tt low} \right)\frac{T_{\tt BNS}}{N_{\tt BNS}} \, ,
\end{align}
with $T_{\tt BNS}$ and $N_{\tt BNS}$ given by Eqs.~(\ref{eq:fitL}) and~(\ref{eq:fitNrb}). 
This speedup is shown in Fig.~\ref{fig:Nbasis_speedup} (bottom), with a reduction in computational cost and time of  $\sim 30$ for the initial detectors (with a cutoff of $f_{\tt low} = 40\,$Hz) and $\sim 150$ once the advanced detectors reach $f_{\tt low} \sim 10\,$Hz. 

{\scriptsize
\begin{algorithm}[H]
\caption{Computing the ROQ integration weights} 
\label{alg:Weights}
\begin{algorithmic}[1]
\vspace{0.2cm}
\State {\bf Input:} $d, S_n, \{ B_j \}_{j=1}^{N}, \Delta f, t_{\tt trigger}, W, \Delta t_c$.
\vskip 10pt
\State Set $n_c=  {\tt int} \left( \left( 2 W \right)/\Delta t_c \right) + 1$ 
\For{$j = 1 \to n_c$} 
\State $T_j = t_{\tt trigger} - W + \left(j - 1 \right) \Delta t_c$
\For{$k = 1 \to N$}
\State Compute $\omega_k(T_j)$ via Eq.~\eqref{eq:wk}
\EndFor
\EndFor
\vskip 10pt
\State {\bf Output:} $\{ T_j \}_{j=1}^{n_c}, \{\{ \omega_k(T_j) \}_{k=1}^N\}_{j=1}^{n_c}$.
\end{algorithmic}
\vspace{0.2cm}
\end{algorithm}
}

\textit{Implementation and numerical studies--} 
We have implemented compressed likelihood evaluations and Algorithm~(\ref{alg:Weights}) in the LAL parameter estimation pipeline, known as LALInference \cite{LAL_software, 2013PhRvD..88f2001A}, naming the resulting variation LALInference$\_$ROQ.

Next we compare MCMC parameter estimation results using the standard version of LALInference to ROQ accelerated studies using LALInference$\_$ROQ. We consider gravitational waveforms emitted from binary neutron star systems with TaylorF2 as the waveform model. We inject synthetic signals embedded in simulated Gaussian noise into the LAL pipeline, for settings anticipating the initial configuration of aLIGO, which should be online within the next two years, using the zero detuned high power PSD \cite{techrep:aLIGOsensitivity}, and the initial configuration of AdV, using Eqn. (6) of \cite{2012arXiv1202.4031M}, with in both cases $f_{low}=40$ Hz.

We take $W= 0.1s$ as the typical time window for the coalescence time $t_c$ of a binary neutron star signal centered around the trigger time~\cite{Smith:2012du, LVCpe:2013}. Following the procedure discussed above, LALInference$\_$ROQ discretizes this prior into $n_c=2,000$ sub-intervals, each of size $\Delta t_c=10^{-5}$s, for which it constructs a unique set of ROQ weights on each sub-interval. A width of $10^{-5}$s ensures that this discretization error is below the measurement uncertainty on the coalescence time, which is typically $\sim 10^{-3}$s \cite{2013PhRvD..88f2001A}. 

\begin{figure}
\includegraphics[width=0.98\linewidth]{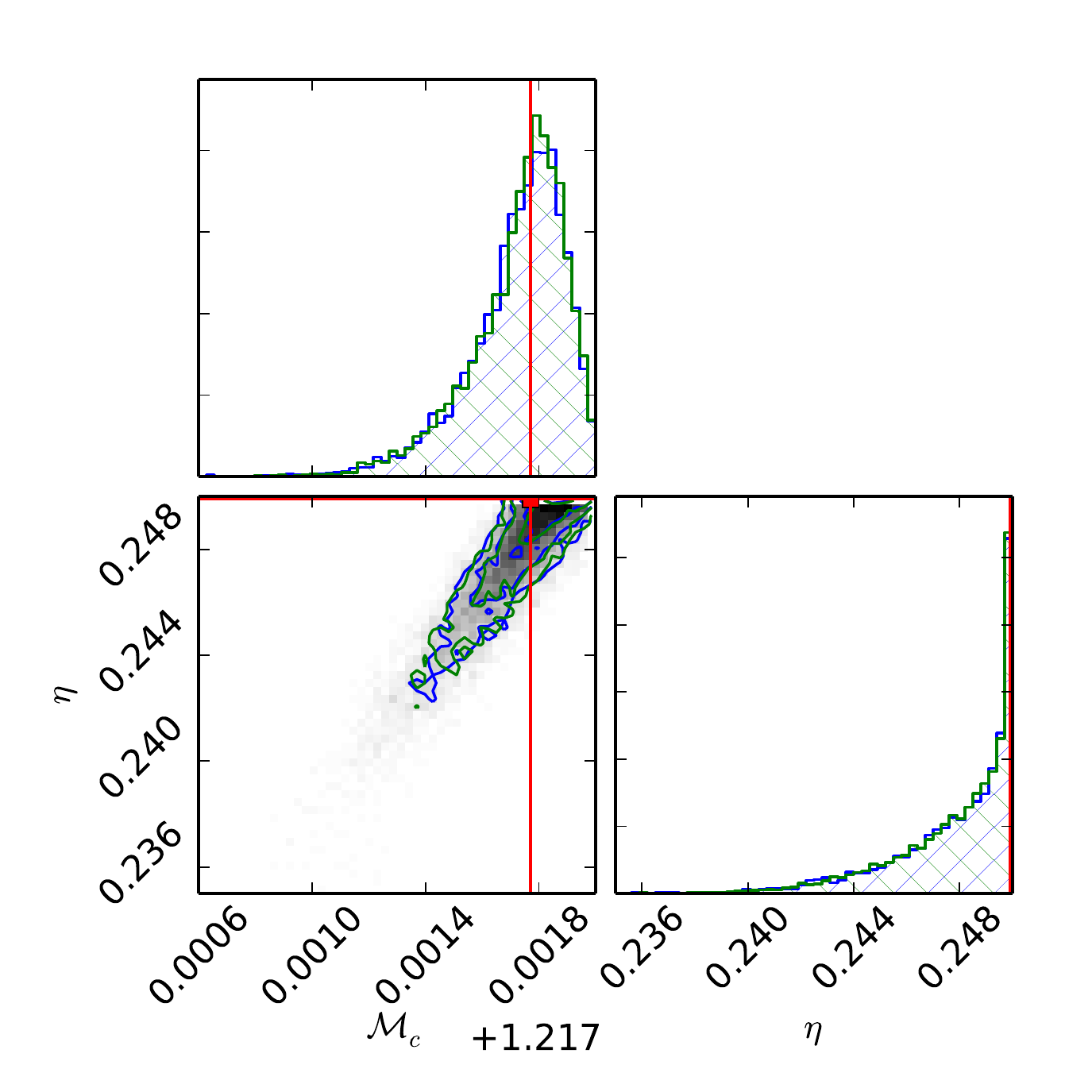}
\caption{Probability density function for the chirp mass $\mathcal{M}_c$ and symmetric mass ratio $\eta$ of a simulated event in LIGO/Virgo data. In green as obtained in $\sim 30$ hours by the standard likelihood, and in blue as obtained in $1$ hour with the ROQ. The injection values are in red, and are listed in Table~\ref{tab:parameters}. The overlap region of the sets of PDFs is the hatched region. Plotting based on \cite{Foreman-Mackey:11020}}
\label{fig:MCMC}
\end{figure}

We found that, as expected, the ROQ and standard likelihood approaches 
produce statistically indistinguishable results for posterior probability density functions over the full $9$-dimensional parameter space. Figure~\ref{fig:MCMC} and Table~\ref{tab:parameters} describe results for the intrinsic mass parameters obtained in one particular MCMC simulation; other simulations were qualitatively similar. It is also useful to quantify the fractional difference in the $9$D likelihood function computed using ROQs and the standard approach. We have observed this fractional error to be
$$
\Delta\log\,\mathcal{L} = 1 - \left( \frac{ \log\,\mathcal{L}}{\log\,\mathcal{L}_{\tt{ROQ}} } \right) \lesssim 10^{-6}
$$ 
in all cases. That is, both approaches are indistinguishable for all practical purposes. 

\begin{table}
\begin{tabular}{ l l l l l l}
  \hline
   & $\mathcal{M}_c\,(M_{\odot})$ & $\eta$ & $m_1\,(M_{\odot})$ & $m_2\,(M_{\odot})$ & SNR \\
  \hline
  injection & 1.2188 & 0.25 & 1.4 & 1.4 & 11.4 \\
  standard & $1.2188^{1.2189}_{1.2184}$  & $0.249^{0.250}_{0.243}$  &  $1.52^{1.66}_{1.41}$  & $1.30^{1.39}_{1.18}$ & 12.9 \\
  ROQ & $1.2188^{1.2189}_{1.2184}$  & $0.249^{0.250}_{0.243}$  &  $1.52^{1.66}_{1.41}$  & $1.30^{1.39}_{1.19}$ & 12.9 \\
  \hline
\end{tabular}
\caption{
Chirp mass $\mathcal{M}_c$, symmetric mass ratio $\eta$, component masses $m_1$ and $m_2$, and Signal-to-Noise Ratio (SNR) of the analysis from Figure~\ref{fig:MCMC}.  
Median value and 90\% credible intervals are provided for both the  standard likelihood (second line) and the ROQ compressed likelihood (third line). 
The SNR is empirically measured from $\mathrm{Likelihood_{max}} \approx \mathrm{SNR}^2 / 2$. The differences between the two methods are dominated by statistics from computing intervals with a finite number of samples. In our analysis, the masses are subject to the constraint $m_1 < m_2$, leading to the true values (where $m_1 = m_2$) being at the edge of the confidence interval.}
\label{tab:parameters}
\end{table}

In addition to providing indistinguishable results, ROQ accelerated inference is significantly faster: The ROQ-based MCMC study with the discussed settings takes $\sim 1$ hour, compared to $\sim 30$ hours using the standard likelihood approach, in remarkable agreement with the expected savings based on operation counts. The wall-time of the analysis is proportional to the total number of posterior samples of the MCMC simulation, which in this case was $\sim 10^{7}$. The startup stage required to build the ROQ weights has negligible cost and is completed in near real-time, $\sim30$s, which is equivalent to $\sim0.028 \%$ of the total cost of a standard likelihood parameter estimation study.

For a lower cutoff frequency of $f_{\tt low} \sim 20$Hz, the speedup reduction is from a couple of weeks to hours.  Once the advanced detectors have achieved their target sensitivity, with $f_{\tt low} \sim 10$Hz, the longest BNS signals will last around $2048$s in duration. Assuming a fiducial high frequency cut-off of $1024\,$Hz, which is approaching the upper limit of where aLIGO/AdV will be sensitive, we estimate datasets as large as $L\sim 1024\text{Hz}^{-1}\times2048\text{s} \sim 10^6$. Assuming that the advanced detectors will require at least $\sim 10^7$ posterior samples, this implies runtimes upwards of $\sim 100$ days and one Petabyte worth of model evaluations using the standard  approach. The results of this Letter indicate that an ROQ approach will reduce this to less than a day. Remarkably, this approach when applied to the advanced detectors operating at design sensitivity will be faster than even the standard likelihood one used for the initial detectors. Additionally, with parallelization of the sum in each likelihood evaluation essentially real-time full Bayesian analysis could be achieved.  More details can be found at \cite{LAL-ROQ}. 

\textit{Outlook--} 
For detectors operating at $f_{\tt low}=40$Hz, around three weeks of real (wall) time are needed to perform a precessing-spin parameter estimation study on a single data stream with standard likelihood evaluations \cite{2013PhRvD..88f2001A}. With a cutoff of $f_{\tt low}=10$Hz, these analyses could take months to years, so techniques for accelerated inference on these models, such as the one presented in this Letter, are expected to play a central role in gravitational-wave astronomy for extracting the full science potential of the upcoming advanced gravitational-wave detectors.

In this Letter we have addressed the issue of fast likelihood evaluation for non-spinning binary neutron star inspirals. Results of previous work indicate that 
significant computational savings are to be expected for other waveform models. For example, in Ref.~\cite{PhysRevD.86.084046} it was found that the number of reduced basis waveforms barely changes as spins are included -- at least in the non-precessing case -- and, by construction, neither does the number of ROQ evaluations. While waveform evaluation is the dominant cost for models which incorporate precession, recent results~\cite{PhysRevLett.113.021101} have shown that ultra-compact bases can also be constructed for fully precessing systems. This might provide a means for constructing fast to evaluate surrogate models of precessing binary inspirals~\cite{Field:2013cfa,Purrer:2014fza}. 

\textit{Acknowledgments--} 
We thank Collin Capano, Peter Diener, Will Farr, Chad Galley, Larry Price, Jorge Pullin, Leo Singer, Alan Weinstein and the LSC CBC and parameter estimation groups for many useful discussions and encouragement throughout this project. PC's work was supported by a Marie Curie Intra-European Fellowship within the 7th European Community Framework Programme (PIEF-GA-2011-299190). SEF thanks hospitality at and financial support from the Institute of Astronomy at Cambridge, UK, where part of this work was done. JG's work was supported by the Royal Society and VR by the LIGO Laboratory and the California Institute of Technology (Caltech). This work was supported in part by NSF grant PHY-1208861 to the University of Maryland (UMD), NSF grant PHY-1500818 to the University of California at San Diego, NSF Grants PHY-1306125 and AST-1333129 to Cornell University, and the Sherman Fairchild Foundation. The authors also gratefully acknowledge the support of the United States National Science Foundation for the construction and operation of the LIGO Laboratory under cooperative agreement NSF-PHY-0757058. This Letter carries LIGO Document Number LIGO-P1400038. Some of the computations were carried out at the Center for Scientific Computation and Mathematical Modeling cluster at UMD and the LIGO Laboratory computer cluster at Caltech. Portions of this research were conducted with high performance computing resources provided by Louisiana State University (http://www.hpc.lsu.edu).

\bibliography{../bibtex-references/references}

\begin{thebibliography}{31}%
\makeatletter
\providecommand \@ifxundefined [1]{%
 \@ifx{#1\undefined}
}%
\providecommand \@ifnum [1]{%
 \ifnum #1\expandafter \@firstoftwo
 \else \expandafter \@secondoftwo
 \fi
}%
\providecommand \@ifx [1]{%
 \ifx #1\expandafter \@firstoftwo
 \else \expandafter \@secondoftwo
 \fi
}%
\providecommand \natexlab [1]{#1}%
\providecommand \enquote  [1]{``#1''}%
\providecommand \bibnamefont  [1]{#1}%
\providecommand \bibfnamefont [1]{#1}%
\providecommand \citenamefont [1]{#1}%
\providecommand \href@noop [0]{\@secondoftwo}%
\providecommand \href [0]{\begingroup \@sanitize@url \@href}%
\providecommand \@href[1]{\@@startlink{#1}\@@href}%
\providecommand \@@href[1]{\endgroup#1\@@endlink}%
\providecommand \@sanitize@url [0]{\catcode `\\12\catcode `\$12\catcode
  `\&12\catcode `\#12\catcode `\^12\catcode `\_12\catcode `\%12\relax}%
\providecommand \@@startlink[1]{}%
\providecommand \@@endlink[0]{}%
\providecommand \url  [0]{\begingroup\@sanitize@url \@url }%
\providecommand \@url [1]{\endgroup\@href {#1}{\urlprefix }}%
\providecommand \urlprefix  [0]{URL }%
\providecommand \Eprint [0]{\href }%
\providecommand \doibase [0]{http://dx.doi.org/}%
\providecommand \selectlanguage [0]{\@gobble}%
\providecommand \bibinfo  [0]{\@secondoftwo}%
\providecommand \bibfield  [0]{\@secondoftwo}%
\providecommand \translation [1]{[#1]}%
\providecommand \BibitemOpen [0]{}%
\providecommand \bibitemStop [0]{}%
\providecommand \bibitemNoStop [0]{.\EOS\space}%
\providecommand \EOS [0]{\spacefactor3000\relax}%
\providecommand \BibitemShut  [1]{\csname bibitem#1\endcsname}%
\let\auto@bib@innerbib\@empty
\bibitem [{\citenamefont {Harry}\ and\ \citenamefont {the LIGO
  Scientific~Collaboration}(2010)}]{Harry2010}%
  \BibitemOpen
  \bibfield  {author} {\bibinfo {author} {\bibfnamefont {Gregory~M}\
  \bibnamefont {Harry}}\ and\ \bibinfo {author} {\bibnamefont {the LIGO
  Scientific~Collaboration}},\ }\bibfield  {title} {\enquote {\bibinfo {title}
  {Advanced ligo: the next generation of gravitational wave detectors},}\
  }\href {http://stacks.iop.org/0264-9381/27/i=8/a=084006} {\bibfield
  {journal} {\bibinfo  {journal} {Classical and Quantum Gravity}\ }\textbf
  {\bibinfo {volume} {27}},\ \bibinfo {pages} {084006} (\bibinfo {year}
  {2010})}\BibitemShut {NoStop}%
\bibitem [{\citenamefont {{The Virgo Collaboration}}(2009)}]{aVirgo}%
  \BibitemOpen
  \bibfield  {author} {\bibinfo {author} {\bibnamefont {{The Virgo
  Collaboration}}},\ }\href
  {https://tds.ego-gw.it/itf/tds/file.php?callFile=VIR-0027A-09.pdf} {\emph
  {\bibinfo {title} {{Advanced Virgo Baseline Design note VIR027A0}}}},\
  \bibinfo {type} {Tech. Rep.}\ (\bibinfo {year} {2009})\BibitemShut {NoStop}%
\bibitem [{\citenamefont {Abadie}\ \emph {et~al.}(2010)\citenamefont {Abadie}
  \emph {et~al.}}]{Abadie:2010cf}%
  \BibitemOpen
  \bibfield  {author} {\bibinfo {author} {\bibfnamefont {J.}~\bibnamefont
  {Abadie}} \emph {et~al.} (\bibinfo {collaboration} {LIGO Scientific
  Collaboration, Virgo Collaboration}),\ }\bibfield  {title} {\enquote
  {\bibinfo {title} {{Predictions for the Rates of Compact Binary Coalescences
  Observable by Ground-based Gravitational-wave Detectors}},}\ }\href {\doibase
  10.1088/0264-9381/27/17/173001} {\bibfield  {journal} {\bibinfo  {journal}
  {Class.Quant.Grav.}\ }\textbf {\bibinfo {volume} {27}},\ \bibinfo {pages}
  {173001} (\bibinfo {year} {2010})},\ \Eprint {http://arxiv.org/abs/1003.2480}
  {arXiv:1003.2480 [astro-ph.HE]} \BibitemShut {NoStop}%
\bibitem [{\citenamefont {Raymond}\ \emph {et~al.}(2009)\citenamefont
  {Raymond}, \citenamefont {van~der Sluys}, \citenamefont {Mandel},
  \citenamefont {Kalogera}, \citenamefont {Rover},\ and\ \citenamefont
  {Christensen}}]{Raymond:2009}%
  \BibitemOpen
  \bibfield  {author} {\bibinfo {author} {\bibfnamefont {V.}~\bibnamefont
  {Raymond}}, \bibinfo {author} {\bibfnamefont {M.~V.}\ \bibnamefont {van~der
  Sluys}}, \bibinfo {author} {\bibfnamefont {I.}~\bibnamefont {Mandel}},
  \bibinfo {author} {\bibfnamefont {V.}~\bibnamefont {Kalogera}}, \bibinfo
  {author} {\bibfnamefont {C.}~\bibnamefont {Rover}}, \ and\ \bibinfo {author}
  {\bibfnamefont {N.}~\bibnamefont {Christensen}},\ }\bibfield  {title}
  {\enquote {\bibinfo {title} {{Degeneracies in Sky Localisation Determination
  from a Spinning Coalescing Binary through Gravitational Wave Observations: a
  Markov-Chain Monte-Carlo Analysis for two Detectors}},}\ }\href@noop {}
  {\bibfield  {journal} {\bibinfo  {journal} {Class. Quant. Grav.}\ }\textbf
  {\bibinfo {volume} {26}},\ \bibinfo {pages} {114007} (\bibinfo {year}
  {2009})}\BibitemShut {NoStop}%
\bibitem [{\citenamefont {Smith}\ \emph {et~al.}(2013)\citenamefont {Smith},
  \citenamefont {Cannon}, \citenamefont {Hanna}, \citenamefont {Keppel},\ and\
  \citenamefont {Mandel}}]{Smith:2012du}%
  \BibitemOpen
  \bibfield  {author} {\bibinfo {author} {\bibfnamefont {R.~J.~E.}\
  \bibnamefont {Smith}}, \bibinfo {author} {\bibfnamefont {K.}~\bibnamefont
  {Cannon}}, \bibinfo {author} {\bibfnamefont {C.}~\bibnamefont {Hanna}},
  \bibinfo {author} {\bibfnamefont {D.}~\bibnamefont {Keppel}}, \ and\ \bibinfo
  {author} {\bibfnamefont {I.}~\bibnamefont {Mandel}},\ }\bibfield  {title}
  {\enquote {\bibinfo {title} {Towards rapid parameter estimation on
  gravitational waves from compact binaries using interpolated waveforms},}\
  }\href {\doibase 10.1103/PhysRevD.87.122002} {\bibfield  {journal} {\bibinfo
  {journal} {Phys. Rev. D}\ }\textbf {\bibinfo {volume} {87}},\ \bibinfo
  {pages} {122002} (\bibinfo {year} {2013})}\BibitemShut {NoStop}%
\bibitem [{\citenamefont {Aasi}\ \emph
  {et~al.}(2013{\natexlab{a}})\citenamefont {Aasi} \emph
  {et~al.}}]{LVCpe:2013}%
  \BibitemOpen
  \bibfield  {author} {\bibinfo {author} {\bibfnamefont {J.}~\bibnamefont
  {Aasi}} \emph {et~al.} (\bibinfo {collaboration} {LIGO-Virgo Scientific
  Collaboration}),\ }\bibfield  {title} {\enquote {\bibinfo {title} {Parameter
  estimation for compact binary coalescence signals with the first generation
  gravitational-wave detector network},}\ }\href {\doibase
  10.1103/PhysRevD.88.062001} {\bibfield  {journal} {\bibinfo  {journal} {Phys.
  Rev. D}\ }\textbf {\bibinfo {volume} {88}},\ \bibinfo {pages} {062001}
  (\bibinfo {year} {2013}{\natexlab{a}})}\BibitemShut {NoStop}%
\bibitem [{\citenamefont {Skilling}(2004)}]{Skilling:2004}%
  \BibitemOpen
  \bibfield  {author} {\bibinfo {author} {\bibfnamefont {J.}~\bibnamefont
  {Skilling}},\ }\bibfield  {title} {\enquote {\bibinfo {title} {Nested
  sampling},}\ }\href@noop {} {\bibfield  {journal} {\bibinfo  {journal} {AIP
  Conf. Ser. Vol. 735}\ ,\ \bibinfo {pages} {395}} (\bibinfo {year}
  {2004})}\BibitemShut {NoStop}%
\bibitem [{\citenamefont {Field}\ \emph {et~al.}(2014)\citenamefont {Field},
  \citenamefont {Galley}, \citenamefont {Hesthaven}, \citenamefont {Kaye},\
  and\ \citenamefont {Tiglio}}]{Field:2013cfa}%
  \BibitemOpen
  \bibfield  {author} {\bibinfo {author} {\bibfnamefont {Scott~E.}\
  \bibnamefont {Field}}, \bibinfo {author} {\bibfnamefont {Chad~R.}\
  \bibnamefont {Galley}}, \bibinfo {author} {\bibfnamefont {Jan~S.}\
  \bibnamefont {Hesthaven}}, \bibinfo {author} {\bibfnamefont {Jason}\
  \bibnamefont {Kaye}}, \ and\ \bibinfo {author} {\bibfnamefont {Manuel}\
  \bibnamefont {Tiglio}},\ }\bibfield  {title} {\enquote {\bibinfo {title}
  {{Fast prediction and evaluation of gravitational waveforms using surrogate
  models}},}\ }\href {\doibase 10.1103/PhysRevX.4.031006} {\bibfield  {journal}
  {\bibinfo  {journal} {Phys. Rev. X}\ }\textbf {\bibinfo {volume} {4}},\
  \bibinfo {pages} {031006} (\bibinfo {year} {2014})},\ \Eprint
  {http://arxiv.org/abs/1308.3565} {arXiv:1308.3565 [gr-qc]} \BibitemShut
  {NoStop}%
\bibitem [{\citenamefont {P{\"u}rrer}(2014)}]{Purrer:2014fza}%
  \BibitemOpen
  \bibfield  {author} {\bibinfo {author} {\bibfnamefont {Michael}\ \bibnamefont
  {P{\"u}rrer}},\ }\bibfield  {title} {\enquote {\bibinfo {title}
  {Frequency-domain reduced order models for gravitational waves from
  aligned-spin compact binaries},}\ }\href
  {http://stacks.iop.org/0264-9381/31/i=19/a=195010} {\bibfield  {journal}
  {\bibinfo  {journal} {Class. Quantum Grav.}\ }\textbf {\bibinfo {volume}
  {31}},\ \bibinfo {pages} {195010} (\bibinfo {year} {2014})},\ \Eprint
  {http://arxiv.org/abs/1402.4146} {arXiv:1402.4146 [gr-qc]} \BibitemShut
  {NoStop}%
\bibitem [{\citenamefont {LAL}()}]{LAL_software}%
  \BibitemOpen
  \bibfield  {author} {\bibinfo {author} {\bibnamefont {LAL}},\ }\href@noop {}
  {}\bibinfo {note}
  {Https://www.lsc-group.phys.uwm.edu/daswg/\\projects/lalsuite.html}\BibitemShut
  {NoStop}%
\bibitem [{\citenamefont {{LIGO Scientific Collaboration}}\ \emph
  {et~al.}(2013)\citenamefont {{LIGO Scientific Collaboration}}, \citenamefont
  {{Virgo Collaboration}}, \citenamefont {{Aasi}}, \citenamefont {{Abadie}},
  \citenamefont {{Abbott}}, \citenamefont {{Abbott}}, \citenamefont {{Abbott}},
  \citenamefont {{Abernathy}}, \citenamefont {{Accadia}}, \citenamefont
  {{Acernese}},\ and\ \citenamefont {et~al.}}]{2013arXiv1304.0670L}%
  \BibitemOpen
  \bibfield  {author} {\bibinfo {author} {\bibnamefont {{LIGO Scientific
  Collaboration}}}, \bibinfo {author} {\bibnamefont {{Virgo Collaboration}}},
  \bibinfo {author} {\bibfnamefont {J.}~\bibnamefont {{Aasi}}}, \bibinfo
  {author} {\bibfnamefont {J.}~\bibnamefont {{Abadie}}}, \bibinfo {author}
  {\bibfnamefont {B.~P.}\ \bibnamefont {{Abbott}}}, \bibinfo {author}
  {\bibfnamefont {R.}~\bibnamefont {{Abbott}}}, \bibinfo {author}
  {\bibfnamefont {T.~D.}\ \bibnamefont {{Abbott}}}, \bibinfo {author}
  {\bibfnamefont {M.}~\bibnamefont {{Abernathy}}}, \bibinfo {author}
  {\bibfnamefont {T.}~\bibnamefont {{Accadia}}}, \bibinfo {author}
  {\bibfnamefont {F.}~\bibnamefont {{Acernese}}}, \ and\ \bibinfo {author}
  {\bibnamefont {et~al.}},\ }\bibfield  {title} {\enquote {\bibinfo {title}
  {{Prospects for Localization of Gravitational Wave Transients by the Advanced
  LIGO and Advanced Virgo Observatories}},}\ }\href@noop {} {\bibfield
  {journal} {\bibinfo  {journal} {ArXiv e-prints}\ } (\bibinfo {year}
  {2013})},\ \Eprint {http://arxiv.org/abs/1304.0670} {arXiv:1304.0670 [gr-qc]}
  \BibitemShut {NoStop}%
\bibitem [{\citenamefont {Antil}\ \emph {et~al.}(2013)\citenamefont {Antil},
  \citenamefont {Field}, \citenamefont {Herrmann}, \citenamefont {Nochetto},\
  and\ \citenamefont {Tiglio}}]{antil2012two}%
  \BibitemOpen
  \bibfield  {author} {\bibinfo {author} {\bibfnamefont {Harbir}\ \bibnamefont
  {Antil}}, \bibinfo {author} {\bibfnamefont {ScottE.}\ \bibnamefont {Field}},
  \bibinfo {author} {\bibfnamefont {Frank}\ \bibnamefont {Herrmann}}, \bibinfo
  {author} {\bibfnamefont {RicardoH.}\ \bibnamefont {Nochetto}}, \ and\
  \bibinfo {author} {\bibfnamefont {Manuel}\ \bibnamefont {Tiglio}},\
  }\bibfield  {title} {\enquote {\bibinfo {title} {Two-step greedy algorithm
  for reduced order quadratures},}\ }\href {\doibase 10.1007/s10915-013-9722-z}
  {\bibfield  {journal} {\bibinfo  {journal} {Journal of Scientific Computing}\
  }\textbf {\bibinfo {volume} {57}},\ \bibinfo {pages} {604--637} (\bibinfo
  {year} {2013})}\BibitemShut {NoStop}%
\bibitem [{\citenamefont {Canizares}\ \emph {et~al.}(2013)\citenamefont
  {Canizares}, \citenamefont {Field}, \citenamefont {Gair},\ and\ \citenamefont
  {Tiglio}}]{Canizares:2013ywa}%
  \BibitemOpen
  \bibfield  {author} {\bibinfo {author} {\bibfnamefont {Priscilla}\
  \bibnamefont {Canizares}}, \bibinfo {author} {\bibfnamefont {Scott~E.}\
  \bibnamefont {Field}}, \bibinfo {author} {\bibfnamefont {Jonathan~R.}\
  \bibnamefont {Gair}}, \ and\ \bibinfo {author} {\bibfnamefont {Manuel}\
  \bibnamefont {Tiglio}},\ }\bibfield  {title} {\enquote {\bibinfo {title}
  {{Gravitational wave parameter estimation with compressed likelihood
  evaluations}},}\ }\href {\doibase 10.1103/PhysRevD.87.124005} {\bibfield
  {journal} {\bibinfo  {journal} {Phys. Rev.}\ }\textbf {\bibinfo {volume}
  {D87}},\ \bibinfo {pages} {124005} (\bibinfo {year} {2013})},\ \Eprint
  {http://arxiv.org/abs/1304.0462} {arXiv:1304.0462 [gr-qc]} \BibitemShut
  {NoStop}%
\bibitem [{\citenamefont {Barrault}\ \emph {et~al.}(2004)\citenamefont
  {Barrault}, \citenamefont {Maday}, \citenamefont {Nguyen},\ and\
  \citenamefont {Patera}}]{Barrault2004667}%
  \BibitemOpen
  \bibfield  {author} {\bibinfo {author} {\bibfnamefont {Maxime}\ \bibnamefont
  {Barrault}}, \bibinfo {author} {\bibfnamefont {Yvon}\ \bibnamefont {Maday}},
  \bibinfo {author} {\bibfnamefont {Ngoc~Cuong}\ \bibnamefont {Nguyen}}, \ and\
  \bibinfo {author} {\bibfnamefont {Anthony~T.}\ \bibnamefont {Patera}},\
  }\bibfield  {title} {\enquote {\bibinfo {title} {An ‘empirical
  interpolation’ method: application to efficient reduced-basis
  discretization of partial differential equations},}\ }\href {\doibase
  http://dx.doi.org/10.1016/j.crma.2004.08.006} {\bibfield  {journal} {\bibinfo
   {journal} {Comptes Rendus Mathematique}\ }\textbf {\bibinfo {volume}
  {339}},\ \bibinfo {pages} {667 -- 672} (\bibinfo {year} {2004})}\BibitemShut
  {NoStop}%
\bibitem [{\citenamefont {Maday}\ \emph {et~al.}(2009)\citenamefont {Maday},
  \citenamefont {Nguyen}, \citenamefont {Patera},\ and\ \citenamefont
  {Pau}}]{Maday_2009}%
  \BibitemOpen
  \bibfield  {author} {\bibinfo {author} {\bibfnamefont {Y.}~\bibnamefont
  {Maday}}, \bibinfo {author} {\bibfnamefont {N.~.C}\ \bibnamefont {Nguyen}},
  \bibinfo {author} {\bibfnamefont {A.~T.}\ \bibnamefont {Patera}}, \ and\
  \bibinfo {author} {\bibfnamefont {S.~H.}\ \bibnamefont {Pau}},\ }\bibfield
  {title} {\enquote {\bibinfo {title} {A general multipurpose interpolation
  procedure: the magic points},}\ }\href {\doibase 10.3934/cpaa.2009.8.383}
  {\bibfield  {journal} {\bibinfo  {journal} {Communications on Pure and
  Applied Analysis}\ }\textbf {\bibinfo {volume} {8}},\ \bibinfo {pages}
  {383--404} (\bibinfo {year} {2009})}\BibitemShut {NoStop}%
\bibitem [{\citenamefont {Chaturantabut}\ and\ \citenamefont
  {Sorensen}(2010)}]{chaturantabut:2737}%
  \BibitemOpen
  \bibfield  {author} {\bibinfo {author} {\bibfnamefont {Saifon}\ \bibnamefont
  {Chaturantabut}}\ and\ \bibinfo {author} {\bibfnamefont {Danny~C.}\
  \bibnamefont {Sorensen}},\ }\bibfield  {title} {\enquote {\bibinfo {title}
  {Nonlinear model reduction via discrete empirical interpolation},}\ }\href
  {\doibase 10.1137/090766498} {\bibfield  {journal} {\bibinfo  {journal} {SIAM
  Journal on Scientific Computing}\ }\textbf {\bibinfo {volume} {32}},\
  \bibinfo {pages} {2737--2764} (\bibinfo {year} {2010})}\BibitemShut {NoStop}%
\bibitem [{\citenamefont {Field}\ \emph {et~al.}(2012)\citenamefont {Field},
  \citenamefont {Galley},\ and\ \citenamefont {Ochsner}}]{PhysRevD.86.084046}%
  \BibitemOpen
  \bibfield  {author} {\bibinfo {author} {\bibfnamefont {Scott~E.}\
  \bibnamefont {Field}}, \bibinfo {author} {\bibfnamefont {Chad~R.}\
  \bibnamefont {Galley}}, \ and\ \bibinfo {author} {\bibfnamefont {Evan}\
  \bibnamefont {Ochsner}},\ }\bibfield  {title} {\enquote {\bibinfo {title}
  {Towards beating the curse of dimensionality for gravitational waves using
  reduced basis},}\ }\href {\doibase 10.1103/PhysRevD.86.084046} {\bibfield
  {journal} {\bibinfo  {journal} {Phys. Rev. D}\ }\textbf {\bibinfo {volume}
  {86}},\ \bibinfo {pages} {084046} (\bibinfo {year} {2012})}\BibitemShut
  {NoStop}%
\bibitem [{\citenamefont {Hoffmann}(1989)}]{Hoffmann_IMGS}%
  \BibitemOpen
  \bibfield  {author} {\bibinfo {author} {\bibfnamefont {Walter}\ \bibnamefont
  {Hoffmann}},\ }\bibfield  {title} {\enquote {\bibinfo {title} {Iterative
  algorithms for gram-schmidt orthogonalization},}\ }\href {\doibase
  10.1007/BF02241222} {\bibfield  {journal} {\bibinfo  {journal} {Computing}\
  }\textbf {\bibinfo {volume} {41}},\ \bibinfo {pages} {335--348} (\bibinfo
  {year} {1989})}\BibitemShut {NoStop}%
\bibitem [{\citenamefont {Allen}\ \emph {et~al.}(2012)\citenamefont {Allen},
  \citenamefont {Anderson}, \citenamefont {Brady}, \citenamefont {Brown},\ and\
  \citenamefont {Creighton}}]{PhysRevD.85.122006}%
  \BibitemOpen
  \bibfield  {author} {\bibinfo {author} {\bibfnamefont {Bruce}\ \bibnamefont
  {Allen}}, \bibinfo {author} {\bibfnamefont {Warren~G.}\ \bibnamefont
  {Anderson}}, \bibinfo {author} {\bibfnamefont {Patrick~R.}\ \bibnamefont
  {Brady}}, \bibinfo {author} {\bibfnamefont {Duncan~A.}\ \bibnamefont
  {Brown}}, \ and\ \bibinfo {author} {\bibfnamefont {Jolien D.~E.}\
  \bibnamefont {Creighton}},\ }\bibfield  {title} {\enquote {\bibinfo {title}
  {Findchirp: An algorithm for detection of gravitational waves from
  inspiraling compact binaries},}\ }\href {\doibase 10.1103/PhysRevD.85.122006}
  {\bibfield  {journal} {\bibinfo  {journal} {Phys. Rev. D}\ }\textbf {\bibinfo
  {volume} {85}},\ \bibinfo {pages} {122006} (\bibinfo {year}
  {2012})}\BibitemShut {NoStop}%
\bibitem [{\citenamefont {Sathyaprakash}\ and\ \citenamefont
  {Schutz}(2009)}]{lrr-2009-2}%
  \BibitemOpen
  \bibfield  {author} {\bibinfo {author} {\bibfnamefont {B.S.}\ \bibnamefont
  {Sathyaprakash}}\ and\ \bibinfo {author} {\bibfnamefont {Bernard~F.}\
  \bibnamefont {Schutz}},\ }\bibfield  {title} {\enquote {\bibinfo {title}
  {Physics, astrophysics and cosmology with gravitational waves},}\ }\href
  {\doibase 10.12942/lrr-2009-2} {\bibfield  {journal} {\bibinfo  {journal}
  {Living Reviews in Relativity}\ }\textbf {\bibinfo {volume} {12}} (\bibinfo
  {year} {2009}),\ 10.12942/lrr-2009-2}\BibitemShut {NoStop}%
\bibitem [{\citenamefont {Buonanno}\ \emph {et~al.}(2009)\citenamefont
  {Buonanno}, \citenamefont {Iyer}, \citenamefont {Ochsner}, \citenamefont
  {Pan},\ and\ \citenamefont {Sathyaprakash}}]{Buonanno:2009zt}%
  \BibitemOpen
  \bibfield  {author} {\bibinfo {author} {\bibfnamefont {Alessandra}\
  \bibnamefont {Buonanno}}, \bibinfo {author} {\bibfnamefont {Bala}\
  \bibnamefont {Iyer}}, \bibinfo {author} {\bibfnamefont {Evan}\ \bibnamefont
  {Ochsner}}, \bibinfo {author} {\bibfnamefont {Yi}~\bibnamefont {Pan}}, \ and\
  \bibinfo {author} {\bibfnamefont {B.S.}\ \bibnamefont {Sathyaprakash}},\
  }\bibfield  {title} {\enquote {\bibinfo {title} {{Comparison of
  post-Newtonian templates for compact binary inspiral signals in
  gravitational-wave detectors}},}\ }\href {\doibase
  10.1103/PhysRevD.80.084043} {\bibfield  {journal} {\bibinfo  {journal}
  {Phys.Rev.}\ }\textbf {\bibinfo {volume} {D80}},\ \bibinfo {pages} {084043}
  (\bibinfo {year} {2009})},\ \Eprint {http://arxiv.org/abs/0907.0700}
  {arXiv:0907.0700 [gr-qc]} \BibitemShut {NoStop}%
\bibitem [{\citenamefont {Singer}\ \emph {et~al.}(2014)\citenamefont {Singer},
  \citenamefont {Price}, \citenamefont {Farr}, \citenamefont {Urban},
  \citenamefont {Pankow} \emph {et~al.}}]{Singer:2014qca}%
  \BibitemOpen
  \bibfield  {author} {\bibinfo {author} {\bibfnamefont {Leo~P.}\ \bibnamefont
  {Singer}}, \bibinfo {author} {\bibfnamefont {Larry~R.}\ \bibnamefont
  {Price}}, \bibinfo {author} {\bibfnamefont {Ben}\ \bibnamefont {Farr}},
  \bibinfo {author} {\bibfnamefont {Alex~L.}\ \bibnamefont {Urban}}, \bibinfo
  {author} {\bibfnamefont {Chris}\ \bibnamefont {Pankow}},  \emph {et~al.},\
  }\bibfield  {title} {\enquote {\bibinfo {title} {{The First Two Years of
  Electromagnetic Follow-Up with Advanced LIGO and Virgo}},}\ }\href@noop {} {\
   (\bibinfo {year} {2014})},\ \Eprint {http://arxiv.org/abs/1404.5623}
  {arXiv:1404.5623 [astro-ph.HE]} \BibitemShut {NoStop}%
\bibitem [{\citenamefont {Eureqa}()}]{eureqa}%
  \BibitemOpen
  \bibfield  {author} {\bibinfo {author} {\bibnamefont {Eureqa}},\ }\href@noop
  {} {}\bibinfo {note} {Http://www.nutonian.com}\BibitemShut {NoStop}%
\bibitem [{\citenamefont {Schmidt}\ and\ \citenamefont
  {Lipson}(2009)}]{Schmidt03042009}%
  \BibitemOpen
  \bibfield  {author} {\bibinfo {author} {\bibfnamefont {Michael}\ \bibnamefont
  {Schmidt}}\ and\ \bibinfo {author} {\bibfnamefont {Hod}\ \bibnamefont
  {Lipson}},\ }\bibfield  {title} {\enquote {\bibinfo {title} {Distilling
  free-form natural laws from experimental data},}\ }\href {\doibase
  10.1126/science.1165893} {\bibfield  {journal} {\bibinfo  {journal}
  {Science}\ }\textbf {\bibinfo {volume} {324}},\ \bibinfo {pages} {81--85}
  (\bibinfo {year} {2009})},\ \Eprint
  {http://arxiv.org/abs/http://www.sciencemag.org/content/324/5923/81.full.pdf}
  {http://www.sciencemag.org/content/324/5923/81.full.pdf} \BibitemShut
  {NoStop}%
\bibitem [{\citenamefont {Mandel}\ \emph {et~al.}(2014)\citenamefont {Mandel},
  \citenamefont {Berry}, \citenamefont {Ohme}, \citenamefont {Fairhurst},\ and\
  \citenamefont {Farr}}]{Mandel:2014tca}%
  \BibitemOpen
  \bibfield  {author} {\bibinfo {author} {\bibfnamefont {Ilya}\ \bibnamefont
  {Mandel}}, \bibinfo {author} {\bibfnamefont {Christopher P~L}\ \bibnamefont
  {Berry}}, \bibinfo {author} {\bibfnamefont {Frank}\ \bibnamefont {Ohme}},
  \bibinfo {author} {\bibfnamefont {Stephen}\ \bibnamefont {Fairhurst}}, \ and\
  \bibinfo {author} {\bibfnamefont {Will~M}\ \bibnamefont {Farr}},\ }\bibfield
  {title} {\enquote {\bibinfo {title} {{Parameter estimation on compact binary
  coalescences with abruptly terminating gravitational waveforms}},}\
  }\href@noop {} {\  (\bibinfo {year} {2014})},\ \Eprint
  {http://arxiv.org/abs/1404.2382} {arXiv:1404.2382 [gr-qc]} \BibitemShut
  {NoStop}%
\bibitem [{\citenamefont {Aasi}\ \emph
  {et~al.}(2013{\natexlab{b}})\citenamefont {Aasi} \emph
  {et~al.}}]{2013PhRvD..88f2001A}%
  \BibitemOpen
  \bibfield  {author} {\bibinfo {author} {\bibfnamefont {J.}~\bibnamefont
  {Aasi}} \emph {et~al.} (\bibinfo {collaboration} {LIGO Collaboration, Virgo
  Collaboration}),\ }\bibfield  {title} {\enquote {\bibinfo {title} {{Parameter
  estimation for compact binary coalescence signals with the first generation
  gravitational-wave detector network}},}\ }\href {\doibase
  10.1103/PhysRevD.88.062001} {\bibfield  {journal} {\bibinfo  {journal}
  {Phys.Rev.}\ }\textbf {\bibinfo {volume} {D88}},\ \bibinfo {pages} {062001}
  (\bibinfo {year} {2013}{\natexlab{b}})},\ \Eprint
  {http://arxiv.org/abs/1304.1775} {arXiv:1304.1775 [gr-qc]} \BibitemShut
  {NoStop}%
\bibitem [{\citenamefont {{LIGO Scientific
  Collaboration}}(2010)}]{techrep:aLIGOsensitivity}%
  \BibitemOpen
  \bibfield  {author} {\bibinfo {author} {\bibnamefont {{LIGO Scientific
  Collaboration}}},\ }\href@noop {} {\emph {\bibinfo {title} {{Advanced LIGO
  anticipated sensitivity curves}}}},\ \bibinfo {type} {Tech. Rep.}\ (\bibinfo
  {year} {2010})\ \bibinfo {note} {{LIGO-T0900288-v3}}\BibitemShut {NoStop}%
\bibitem [{\citenamefont {{Manzotti}}\ and\ \citenamefont
  {{Dietz}}(2012)}]{2012arXiv1202.4031M}%
  \BibitemOpen
  \bibfield  {author} {\bibinfo {author} {\bibfnamefont {A.}~\bibnamefont
  {{Manzotti}}}\ and\ \bibinfo {author} {\bibfnamefont {A.}~\bibnamefont
  {{Dietz}}},\ }\bibfield  {title} {\enquote {\bibinfo {title} {{Prospects for
  early localization of gravitational-wave signals from compact binary
  coalescences with advanced detectors}},}\ }\href@noop {} {\bibfield
  {journal} {\bibinfo  {journal} {ArXiv e-prints}\ } (\bibinfo {year}
  {2012})},\ \Eprint {http://arxiv.org/abs/1202.4031} {arXiv:1202.4031 [gr-qc]}
  \BibitemShut {NoStop}%
\bibitem [{\citenamefont {Foreman-Mackey}\ \emph {et~al.}(2014)\citenamefont
  {Foreman-Mackey}, \citenamefont {Price-Whelan}, \citenamefont {Ryan},
  \citenamefont {Emily}, \citenamefont {Smith}, \citenamefont {Barbary},
  \citenamefont {Hogg},\ and\ \citenamefont {Brewer}}]{Foreman-Mackey:11020}%
  \BibitemOpen
  \bibfield  {author} {\bibinfo {author} {\bibfnamefont {Dan}\ \bibnamefont
  {Foreman-Mackey}}, \bibinfo {author} {\bibfnamefont {Adrian}\ \bibnamefont
  {Price-Whelan}}, \bibinfo {author} {\bibfnamefont {Geoffrey}\ \bibnamefont
  {Ryan}}, \bibinfo {author} {\bibnamefont {Emily}}, \bibinfo {author}
  {\bibfnamefont {Michael}\ \bibnamefont {Smith}}, \bibinfo {author}
  {\bibfnamefont {Kyle}\ \bibnamefont {Barbary}}, \bibinfo {author}
  {\bibfnamefont {David~W.}\ \bibnamefont {Hogg}}, \ and\ \bibinfo {author}
  {\bibfnamefont {Brendon~J.}\ \bibnamefont {Brewer}},\ }\bibfield  {title}
  {\enquote {\bibinfo {title} {{triangle.py v0.1.1}},}\ }\href {\doibase
  {10.5281/zenodo.11020}} {\  (\bibinfo {year} {2014}),\
  {10.5281/zenodo.11020}}\BibitemShut {NoStop}%
\bibitem [{LAL-ROQ()}]{LAL-ROQ}%
  \BibitemOpen
  LAL-ROQ,\ \href@noop {} {}\bibinfo {note}
  {\url{http://www.ligo.caltech.edu/~vraymond/rom/20Hz/}}\BibitemShut {NoStop}%
\bibitem [{\citenamefont {Blackman}\ \emph {et~al.}(2014)\citenamefont
  {Blackman}, \citenamefont {Szilagyi}, \citenamefont {Galley},\ and\
  \citenamefont {Tiglio}}]{PhysRevLett.113.021101}%
  \BibitemOpen
  \bibfield  {author} {\bibinfo {author} {\bibfnamefont {Jonathan}\
  \bibnamefont {Blackman}}, \bibinfo {author} {\bibfnamefont {Bela}\
  \bibnamefont {Szilagyi}}, \bibinfo {author} {\bibfnamefont {Chad~R.}\
  \bibnamefont {Galley}}, \ and\ \bibinfo {author} {\bibfnamefont {Manuel}\
  \bibnamefont {Tiglio}},\ }\bibfield  {title} {\enquote {\bibinfo {title}
  {Sparse representations of gravitational waves from precessing compact
  binaries},}\ }\href {\doibase 10.1103/PhysRevLett.113.021101} {\bibfield
  {journal} {\bibinfo  {journal} {Phys. Rev. Lett.}\ }\textbf {\bibinfo
  {volume} {113}},\ \bibinfo {pages} {021101} (\bibinfo {year}
  {2014})}\BibitemShut {NoStop}%
\end{thebibliography}%

\end{document}